\documentclass[12pt]{article}
\usepackage{amsfonts}
\usepackage{}
\usepackage{bbm}
\usepackage{bm}
\usepackage{mathrsfs}
\usepackage{geometry}             
\usepackage{graphicx}
\usepackage{amssymb}
\usepackage{amsmath}
\usepackage{epstopdf}

\usepackage{comment}

\usepackage[hyperindex=true,
          pdfstartview=FitH,
          bookmarksnumbered=true,
          bookmarksopen=true,
          citecolor=blue,
          linkcolor=blue,
          colorlinks=true,
          pdfborder=001,
          unicode]{hyperref}

\newcommand{\rd}[1]{\mathrm{d}#1}

\newcommand{\re}{\mathrm{e}}
\newcommand{\la}{\langle}
\newcommand{\ra}{\rangle}

\usepackage{xcolor}  
\definecolor{mygreen}{RGB}{44,85,17}
\definecolor{myblue}{RGB}{34,31,217}
\definecolor{mybrown}{RGB}{194,164,113}
\definecolor{myred}{RGB}{255,66,56}
\definecolor{mypurple}{RGB}{200,36,176}

\begin{document}

\title{Relativistic transformation of thermodynamic parameters and refined Saha equation}
\author{Xin Hao${}^1$, Shaofan Liu${}^2$
and Liu Zhao${}^2$\thanks{Correspondence author.}
\vspace{-5pt}\\
\small ${}^1$College of Physics, Hebei Normal University, Shijiazhuang 050024, China\\
\small ${}^2$School of Physics, Nankai University, Tianjin 300071, China\\
{\em email}: \href{mailto:xhao@hebtu.edu.cn}
{xhao@hebtu.edu.cn},
\href{mailto:2120190132@mail.nankai.edu.cn}
{2120190132@mail.nankai.edu.cn} \\
and
\href{mailto:lzhao@nankai.edu.cn}{lzhao@nankai.edu.cn}
}
\date{}
\maketitle

\begin{abstract}
The relativistic transformation rule for temperature is a subject
under debate for more than 110 years. Several incompatible proposals
exist in the literature, but a final resolution is still missing.
In this work, we reconsider the problem of relativistic transformation
rules for a number of thermodynamic parameters, including temperature,
chemical potential, pressure, entropy and enthalpy densities for
a relativistic perfect fluid using relativistic kinetic theory.
The analysis is carried out in a fully relativistic covariant manner,
and the explicit transformation rules for the above quantities
are obtained in both Minkowski and Rindler spacetimes. Our results
suggest that the temperature of a moving fluid appears to be colder,
supporting the proposal by de Broglie, Einstein and Planck in contrast to other
proposals. Moreover, in the case of Rindler fluid, our work indicates that,
the total number of particles and the total entropy of a
perfect fluid in a box whose bottom is parallel to the Rindler horizon
are proportional to the area of the bottom, but are
independent of the height of the box, provided the bottom of the box
is sufficiently close to the Rindler horizon. The area dependence of the
particle number implies that the particles tend to be gathered toward the
bottom of the box and hence implicitly determines the distribution
of chemical potential of the system, whereas the area dependence of
the entropy indicates that the entropy is still additive and may find
some applications in explaining the area law of black hole entropy.
As a by product, we also  obtain a relativistically refined version
of the famous Saha equation which holds in both Minkowski and Rindler spacetimes.

\end{abstract}

\vspace{3em}
\newpage

\section{Introduction}

Of all branches of modern physics, classical thermodynamics and relativity
are outstanding in the sense that they describe the universal rules that
every physical system must obey, irrespective of the detailed matter
contents of the system. There are only two requirements for
classical thermodynamics to hold: i) the physical system needs to be
macroscopic, i.e. containing a large number of microscopic degrees of freedom,
ii) the system needs to be at thermodynamic
equilibrium with uniform temperature and pressure. In contrast, there seems
to be no requirement for any physical system to obey the principles
of relativity, although the special relativistic effects can only become
manifest when the system undergoes very fast motion in comparison to the speed
of light, and the general relativistic effects can only become manifest
when the system contains a huge amount of mass and/or energy.

It has long been fascinating to consider situations when both the
principles of classical thermodynamics and relativity apply. Such situations
involve macroscopic system which either undergoes relativistic motion
or moves in curved spacetime. The endeavors in combining
thermodynamics and relativity have lasted for over 110 years. However,
the outcome is quite controversial.
Even without considering the general relativistic effects,
the combination of classical thermodynamics and special relativity
has led to several contradictory results on the transformation rule
for temperature. Basically, there are four major
views on such transformations, each labeled by the names of the
corresponding researchers below (wherein $\gamma>1$ is the Lorentz factor):

(i) de Broglie \cite{deBroglie1995}, Einstein \cite{Einstein1907} and
Planck \cite{Planck1908}: moving bodies appear cooler, $T' = \gamma^{-1} T$;

(ii) Eddington \cite{Eddington1923}, Ott \cite{Ott1963} and Arzelies \cite{Ar}:
moving bodies appear hotter, $T' = \gamma T$;

(iii) Landsberg \cite{Landsberg1966,Landsberg1967}: temperature is a
relativistic invariant, $T'=T$;

(iv)  Cavalleri, Salgarelli \cite{Cavalleri1969} and Newburgh \cite{Newburgh1979}:
no unique such transformation because thermodynamics is defined only in rest frame.

Quite notably, Einstein seems to have supported each of the four views in his life
\cite{Liu}, and Lansberg turned to the fourth view in his later career
\cite{Ls1, Ls2}.
The debates between all these different views remain open \cite{Farias2017}
and a huge number of papers have been published on the same or related subjects.
It is remarkable that the standing point of the fourth view by Cavalleri {\it et al}
lies in that a system at thermodynamic equilibrium must be static
\cite{Cavalleri1969, Newburgh1979} and hence
excludes the existence of macroscopic flow which is inherently implied by
global relativistic motion of the system, and that a moving observer
in a heat reservoir cannot detect a blackbody spectrum \cite{Ls1, Ls2},
which implies the nonexistence of a uniform temperature.
Such reasonings, however, should not be taken to be sufficient
justifications for the fourth view, because there are situations beyond
thermodynamic equilibrium when one could talk about the temperature,
pressure and entropy \emph{etc.}, at least locally, of a given macroscopic system,
for instance for systems which are not in global but in
local equilibrium, or for systems under detailed balance. For such systems,
the classical equilibrium thermodynamics does not apply, however, a description
using kinetic theory still works well.

Among the existing papers on related subjects
(not necessarily considering the temperature transformations), some
considerations from the point of view of statistical mechanics or
kinetic theory have been introduced. Some authors
\cite{Padmanabhan2011,Padmanabhan2017,Kim2017,Li:2021zuw}
started right from equilibrium statistical mechanics or Gibbs distributions
and the formulations were often not presented in explicitly relativistic
covariant fashion, hence not best suited for analyzing the
transformation rules for macroscopic parameters.
Some other works either dealt with the debates about
the correct relativistic distribution function \cite{Cubero2007,Montakhab2009}
or introduced some modifications to the distribution function
\cite{Kaniadakis2002}.

In order to solve the puzzles on the relativistic transformation rules
for macroscopic parameters, the necessary statistical mechanics tool
needs to be relativistic covariant and applicable to systems out
of thermodynamic equilibrium. Such a theory exists and is known as
the relativistic kinetic theory. It was established almost
right after the first view by de Broglie, Einstein and Planck was
proposed  \cite{Juttner1911}. Therefore, it is
tempting to reconsider the relativistic transformations for macroscopic
parameters like temperature and pressure from the point of view of
relativistic kinetic theory. As far as we know, a fully
covariant treatment for this problem using relativistic kinetic theory has
not been reported before in literature, so we decide to
work it out by ourselves.

Before dwelling into the detailed analysis, it is worth pointing out that
the aforementioned debates stem largely from the way that the question is raised.
All previous works on this subject prescribe the question as follows:
{\em Assuming the
temperature of a system in (local) thermodynamic equilibrium is $T$ in the rest frame.
What is its temperature $T'$ in a frame in which the system undergoes fast motion? }
An alternative prescription of the question which does not rely on
the choice of coordinate frames can be given as follows.
{\em Assuming the system is at (local)
thermodynamic equilibrium at temperature $T$ with respect to the comoving observer.
What is its temperature $T'$ with respect to a non-comoving observer?}
The two prescriptions
differ from each other in the reason why such changes happen. The first
prescription attributes the change of temperature to the change of coordinate frames, while
the second  prescription attributes the change to the change of observers.
Even so, both prescriptions
quest the change of the temperature of the system at same macrostate, and the change
of temperature in both prescriptions arises purely from kinematic effects.
Therefore, both prescriptions can be dubbed as the kinematic version of the question.
There is a dynamic version of the question which quests for the temperature of the
system which is initially at rest and then pushed into fast motion. This version
breaks the initial macrostate and will not be discussed here.

In this work, we will take the second kinematic prescription as the starting point.
The reason to take the second rather than the first prescription is due to the
following considerations. First of all, most thermodynamic parameters have
phenomenological interpretations and their values are naturally observer dependent.
The first kinematic prescription does not provide information about such dependences.
Second, we hope to understand the relativistic transformation rules for
thermodynamic parameters in more generic spacetimes rather than just in Minkowski
spacetime. Therefore, the coordinate changes do not necessarily belong to the set of
Lorentz transformations. Last but not least, we will show that most of the thermodynamic
parameters (or densities thereof) can be defined as scalars with respect to the coordinate
transformations, whereas their transformation rules under change of observers are
still nontrivial. This last reasoning indicates that the first prescription is actually
ill-posed.

As will be shown in the main context, our
analysis indicates that the transformation rule of temperature agrees with the
view of de Broglie, Einstein and Planck, but with the addition of the transformation
rules for a number of other thermodynamic parameters,
notably including the chemical potential $\mu$, the particle number density $n$ and
the enthalpy density $w$. Our analysis indicates that the transformation rules
of those parameters are identical
in both Minkowski and Rindler spacetimes, and we expect that
the same rules should also be valid in other backgrounds as well
due to the fully relativistic covariant formalism. In the case
of Rindler background, we shall also show that
the total number of particles and the total entropy of a
perfect fluid system in a box are proportional to the area of the
bottom of the box which is parallel to the Rindler horizon, but are
independent of the height of the box, provided the bottom of the box
is sufficiently close to the Rindler horizon. Moreover,
since the chemical potential is explicitly calculated in our
considerations, it is straightforward to obtain a relativistically
refined version of the famous Saha equation \cite{saha}
which characterizes the local chemical equilibrium in the
ultra relativistic regime.

\section{Elements of relativistic kinetic theory}

Our main tool is relativistic kinetic theory based on a
covariant generalization of Boltzmann equation. This theory is a subfield
of non-equilibrium statistical physics, and the application of this
theory in our analysis implies some
microscopic considerations are involved.
Different from the Gibbs method for equilibrium system, in kinetic theory,
the distribution function is taken to be the one particle distribution
function (1PDF) $f(x,p)$ which is defined to be the local particle
number density in the one particle phase space. In relativistic settings,
one often enlarges the one particle phase space to the tangent bundle of
the full spacetime\footnote{For the sake of generality, we do not require
the spacetime to be flat, thus the formulation of relativistic kinetic
theory to be described below applies to both special and general
relativistic cases.}, of which the fibre space is spanned by the
proper momentum vector $p^\mu$ for individual particles which obey
the mass shell constraint $p^\mu p_\mu=-m^2 c^2$, where $m$
is the rest mass of the particle. The enlarged one particle phase
space is endowed with a relativistic invariant measure
\cite{Sarbach2013,Sarbach:2013uba}
$\bm{\Omega}=\bm\varpi\wedge \bm\epsilon$, where
$\bm\varpi=\frac{\sqrt{g}}{|p_0|}\rd^{d}p$ is the momentum space volume
element\footnote{The appearance of $\rd^{d}p$ rather than $\rd^{d+1}p$
is due to the mass shell constraint.},
$\bm\epsilon =\sqrt{g} \rd^{d+1} x$ is the spacetime volume element,
$g=|\mathrm{det}(g_{\mu\nu})|$, and $d+1$ is the spacetime dimension.
For a dilute gas system, in much of the region
in phase space, the 1PDF is locally conserved
\begin{align}
\mathscr{L}_{{\mathcal H}} f =0,
\label{Liouville}
\end{align}
where $\mathscr{L}_{{\mathcal H}}$ denotes the Lie derivatives along the
Hamiltonian vector field $\mathcal{H}$.
This is known as the relativistic Liouville equation. However,
taking into account the contribution from the inter-particle
scatterings, the Liouville
equation should be replaced by the Boltzmann equation
\begin{align}
\mathscr{L_{\mathcal H}} f= \mathcal C(x,p),
\end{align}
where the scattering integral $\mathcal C(x,p)$ is a non-local integral
in terms of the 1PDF $f(x,p)$ and the local transition rate
$W(x|p_1,p_2;p_3,p_4)$ if two particle scatterings are dominated.
Assuming that the above equation is solved, all macroscopic evolutions
of the non-equilibrium system will be determined by the 1PDF, including
the particle number current $N^\mu$, the energy-momentum tensor
$T^{\mu\nu}$ and the entropy current $S^\mu$:
\[
N^\mu=c\int\bm\varpi p^\mu f, \qquad
T^{\mu\nu}=c\int\bm\varpi p^\mu p^\nu f, \qquad
S^\mu = -k_{\mathrm{B}} c \int \bm\varpi p^\mu f
\left[\log(h^df) - 1\right],
\]
wherein $h$ is the Planck constant and $k_{\mathrm{B}}$ is the Boltzmann constant.
Since Boltzmann equation is an integro-differential equation,
finding a solution is a highly nontrivial task.
Fortunately, we can draw some interesting conclusions without too
much effort on solving the equation. As long as Boltzmann
equation holds, the particle number is conserved
$\nabla_\mu N^\mu =0$, the hydrodynamic
equation is established $\nabla_\mu T^{\mu\nu} = 0$, and the entropy
never decrease $\mathcal{G} \equiv \nabla_\mu S^\mu \geq 0$.
$\mathcal{G}$ is known as the local entropy production rate, and
when $\mathcal{G} = 0$, the system is in detailed balance.

Under detailed balance, one can show that $\log(h^d f)$ is
additive and must be linearly dependent on additive conserved
quantities of the macroscopic system. Meanwhile, the
Boltzmann equation degenerates into the Liouville equation
\eqref{Liouville}.  With all these conditions we can conclude that
the general form of the 1PDF under detailed balance is
\begin{align}
f_0 = \frac{1}{h^d} \exp(-\alpha+ \mathcal B_\mu p^\mu),
\label{eq1PDF}
\end{align}
where $\alpha$ and $\mathcal B_\mu$ are undetermined coefficients
which satisfy
\begin{align}
-p^\mu \partial_\mu \alpha + p^\mu p^\nu
\nabla_ \mu\mathcal B_\nu = 0.
\end{align}
We assume that the motion of individual particles is completely
random, then the coefficients in front of the linear and
quadratic terms in the momentum must separately be zero.
Hence we have
\begin{align}
\partial_\mu \alpha = 0, \qquad \text{and} \qquad
\nabla_{(\mu}\mathcal B_{\nu)} = 0,
\label{Bmu}
\end{align}
Ignoring the trivial solution $\mathcal B^\mu =0$, it follows
from eq. \eqref{Bmu} that $\mathcal B^\mu$ must be a Killing vector
field. Let us remind that the 1PDF \eqref{eq1PDF} applies to
any classical macroscopic system in any spacetime admitting a Killing
vector field $\mathcal B^\mu$, provided the system is under detailed balance.
The form of the 1PDF \eqref{eq1PDF} reminds us of the famous
J\"uttner distribution \cite{Juttner1911,DeGroot:1980dk,Cer,Hak}
\begin{align}
f=\frac{1}{h^d} \exp(-\alpha+ \beta U_\mu p^\mu)
\label{Jutn}
\end{align}
for relativistic fluid in equilibrium, wherein $U_\mu$ is the average
proper velocity of the fluid element and $\beta$ is the inverse temperature.
However, at this point, we do not attempt to link $\mathcal B_{\mu}$ with $U_\mu$,
and do not introduce {\em a priori} an inverse temperature.
The system described by the 1PDF  \eqref{eq1PDF} is only
in detailed balance but not in thermodynamic equilibrium.
As will be shown later, if we choose the instantaneously comoving observer
while describing the motion of the fluid,
the distribution \eqref{eq1PDF} indeed reduces to the standard
J\"uttner distribution \eqref{Jutn}. However, if we choose an arbitrary
non-comoving observer, eq.\eqref{eq1PDF} will differ from the
J\"uttner distribution \eqref{Jutn}, which will allow us to
uncover the transformation rules for macroscopic quantities.

The solution to eq. \eqref{Bmu} can be non-unique because the
spacetime may admits several independent Killing vector fields.
In such cases, ${\mathcal B}^{\mu}$ can be either timelike or spacelike.
Accordingly, the quantity ${\mathcal B}_{\mu} p^\mu$ can be proportional either
to the single particle energy or to certain momentum component(s)
of the single particle. It is not surprising that the
1PDF can depend on the momentum component(s) of the particle. Even
in non-relativistic statistical mechanics, the distribution function can
depend on the particle's momentum if the system maintains spatial
translation symmetry. However, there may be surprising cases if the spacetime
has no timelike Killing vector but do have several spacelike ones.
In such spacetimes, the detailed balance distribution of the form \eqref{eq1PDF}
can still be achieved. This possibility is of little interest if one considers
the fact that the macroscopic system needs to be spatially confined by a potential
which breaks all spatial translational symmetries. In short, although
the solutions to eq.\eqref{Bmu} depends solely on the spacetime
diffeomorphism symmetry, the physical choice of the solution
can only be determined by considering the symmetry of
the potential as well as the boundary conditions of the system.
In the present work we assume that the spacetime and the boundary conditions of
the system altogether admit at least one timelike $\mathcal B^\mu$.
We also assume that $\mathcal B^\mu$ is
normalized as $\mathcal B^2 = -\beta^2 c^2$, in which the physical
meaning of the scalar parameter $\beta$ is yet to be
interpreted. In order to have an intuitive understanding about this parameter,
let us temporarily consider a two-component mixture consisting of species (I) and (II)
between which only elastic scatterings could occur. For such a system,
the detailed balance condition reduces to the equation \cite{DeGroot:1980dk}
\begin{align*}
f_{\text I}(p_{\text I})\,f_{\text{II}}(p_{\text{II}}) =
f'_{\text I}(p'_{\text I})\,f'_{\text{II}}(p'_{\text{II}}),
\end{align*}
where unprimed and primed symbols represent respectively the corresponding quantities
before and after the elastic scattering. Accordingly
\begin{align*}
\mathcal B_{(\text{I})\mu}p_{({\text{I}})}{}^\mu
+\mathcal B_{(\text{II})\mu}p_{({\text{II}})}{}^\mu
=\mathcal B_{(\text{I})\mu}p'_{({\text{I}})}{}^\mu
+\mathcal B_{(\text{II})\mu}p'_{({\text{II}})}{}^\mu
\end{align*}
is satisfied for any elastic binary collision,
$p_{({\text{I}})}{}^\mu + p_{({\text{II}})}{}^\mu =
p'_{({\text{I}})}{}^\mu + p'_{({\text{II}})}{}^\mu$. Further,
with the same boundary conditions, $\mathcal B_{(\text{I})}^\mu$
and $\mathcal B_{(\text{II})}^\mu$ must be collinear. Therefore
$\beta_{(\text{I})} $ and $\beta_{(\text{II})} $ must be equal,
which means that locally there is a commonly shared scalar for two comoving
systems under detailed balance. In this sense, $\beta$ may be used while defining
temperature. This argument agrees in spirits with Ref. \cite{Cubero2007}.
Later, we will show that $\beta$ is indeed connected to the inverse temperature
observed by comoving observers in more generic settings rather than just in the
two component mixture described above.

Using the detailed balance distribution \eqref{eq1PDF}, we can rewrite the
particle number current, the energy-momentum tensor and the entropy current
under detailed balance as follows,
\begin{align}
N^\mu &= \frac{c\,\re^{-\alpha}}{h^d}\int
\frac{\sqrt g\,\mathrm d^{d}p}{|p_0|} p^\mu \re^{\mathcal B_\rho p^\rho},
\quad T^{\mu\nu} = \frac{c\,\re^{-\alpha}}{h^d}
\int \frac{\sqrt g\,\mathrm d^dp}{|p_0|} p^\mu p^\nu
\re^{\mathcal B_\rho p^\rho},  \label{N&T} \\
S^\mu &= \frac{k_{\mathrm{B}} c\,\re^{-\alpha}}{h^d}
\int \frac{\sqrt g\,\mathrm d^dp}{|p_0|}
\left(1+\alpha - \mathcal B_\nu p^\nu\right)p^\mu
\re^{\mathcal B_\rho p^\rho}
= k_{\mathrm{B}} \left(1+\alpha\right)N^\mu
- k_{\mathrm{B}}\,\mathcal B_\nu T^{\mu\nu}.
\label{Smu}
\end{align}
It is evident that provided the conditions \eqref{Bmu} are satisfied, the
hydrodynamic equations $\nabla_\mu N^\mu=0, \nabla_\mu T^{\mu\nu}=0$ and
the detailed balance condition $\nabla_\mu S^{\mu}=0$ are explicitly
satisfied. Therefore, the macroscopic system is non-dissipative and can be
modeled as a perfect fluid. Notice, however, since $N^{\mu}$, $T^{\mu\nu}$
and $S^{\mu}$ are all tensorial objects, the components of these
objects depend explicitly on the choices of the spacetime geometry and the
coordinate system. According to the principles of relativity, the
characteristic properties of the perfect fluid need to be described in a
way which is independent of the coordinate choice. This can be done
by introducing a number of scalar density parameters. Even so,
an explicit specification for the spacetime metric $g_{\mu\nu}$ is inevitable.
In the next section, we shall calculate the physical quantities for the
perfect fluid in Minkowski spacetime and uncover the relativistic transformation rules
for a number of scalar observables of the perfect fluid.

\section{Transformation rules in Minkowski spacetime} \label{sec:Minkowski}

In this section, we shall fix the spacetime to be Minkowskian, and for
simplicity, we shall set the spacetime dimension to be $3+1$, and
take the coordinates to be cartesian. Under such setting, the Killing
vector field $\mathcal B^\mu$ can be taken as
$\mathcal B^\mu = \beta (\partial_t)^\mu =\left(\beta c,0,0,0\right)$,
the particle number current $N^\mu$ and the energy-momentum tensor
$T^{\mu\nu}$ as given in \eqref{N&T} can be explicitly evaluated to be
\begin{align}
N^\mu &= \left(\frac{4\pi \re^{-\alpha}}{\lambda_{C}^{3} \zeta} c\,K_2(\zeta)
,0,0,0\right),    \label{Ncomp}  \\
T^{\mu \nu} &=
\frac{4\pi \re^{-\alpha}}{\lambda_{C}^{3} \zeta} mc^2 K_{2}(\zeta)
\left(\begin{array}{cccc}
\left[K_{3}(\zeta)/K_2(\zeta)-\zeta^{-1}\right] & 0 & 0 & 0 \\
0 & \zeta^{-1} & 0 & 0 \\
0 & 0 & \zeta^{-1} & 0 \\
0 & 0 & 0 & \zeta^{-1}
\end{array}\right),
\label{Tcomp}
\end{align}
where $\lambda_C = \frac{h}{mc}$ is the Compton wave length of the
constituent particles,
$K_\nu(\zeta)$ is the modified Bessel function of the
second kind, and $\zeta = \beta mc^2$ is a dimensionless parameter.
It follows from eq.\eqref{Ncomp} that the particle number current has
no spatial component, which reflects the fact that we are working in a
comoving coordinate system. The particle
number density can be read off from the temporal component of $N^\mu$,
\begin{align}
\bar n =\frac{1}{c} N^0
= \frac{4\pi \re^{-\alpha}}{\lambda_{C}^{3} \zeta} K_2(\zeta).
\label{nb}
\end{align}
The energy density is the 00 component of
the energy momentum tensor which reads
\begin{align}
\bar \epsilon = T^{00}
= \frac{4\pi \re^{-\alpha}}{\lambda_{C}^{3} \zeta}mc^2
\left[K_{3}(\zeta)-\zeta^{-1}K_{2}(\zeta)\right].
\label{eb}
\end{align}
Moreover, from eq.\eqref{Tcomp}, one can see that the stress tensor
(regarded as the spatial-spatial part of the energy momentum tensor)
is diagonal, $\bar{\mathcal{S}}^{ij} = \bar P \delta^{ij}$, which
means that the perfect fluid is characterized by an isotropic pressure
\begin{align}
\bar P = \bar{T}^{ii}
= \frac{4\pi \re^{-\alpha}}{\lambda_{C}^{3} \zeta^2} mc^2 K_{2}(\zeta)
=\zeta^{-1} mc^2 \bar n =\frac{\bar n}{\beta}.
\label{Pb}
\end{align}

The particle number density $\bar{n}$, the pressure $\bar{P}$ and the
energy density $\bar{\epsilon}$ given above are taken to be some
specific component (or sum of components) of the relevant tensors.
This may raise concerns about the relativistic invariance (or covariance) of
these quantities. However, the fact is that there is an observer,
say $\bar{\mathcal O}$, hidden in the above result. To be specific,
at each given spacetime event, the proper velocity of
$\bar{\mathcal O}$ is $\bar Z^\mu = (\partial_t)^\mu$, and
$\bar{n}$ and $\bar{\epsilon}$ are actually the following
scalar observables measured by the observer
$\bar{\mathcal O}$,
\begin{align}
\bar n = - \frac{1}{c^2} \bar Z_\mu N^\mu,
\qquad \bar \epsilon = \frac{1}{c^2} \bar Z_\mu \bar Z_\nu T^{\mu\nu},
\label{covdefs}
\end{align}
and the pressure $\bar{P}$ is simply one third of the trace of the pure
stress tensor $\bar{\mathcal{S}}^{\mu\nu}$ (now regarded as a tensor
on the full spacetime) measured by $\bar{\mathcal O}$:
\begin{align}
\bar{\mathcal{S}}^{\mu\nu} \equiv T^{\rho\sigma}\bar\Delta_\rho{}^\mu \bar\Delta_\sigma{}^\nu,
\qquad \bar P =
\frac{1}{3} \bar \Delta_{\mu\nu} \bar{\mathcal{S}}^{\mu\nu},
\label{bPstress}
\end{align}
where
\[
\bar \Delta^{\mu\nu}
= \eta^{\mu\nu}+\frac{1}{c^2}\bar Z^\mu \bar Z^\nu
\]
is the normal projection tensor associated to the
observer $\bar{\mathcal{O}}$ which satisfies
\begin{align*}
&\bar \Delta_{\mu\nu} \bar Z^\mu=0,\quad
\bar \Delta_{\mu\nu} v^\mu=v^\nu,\quad \forall v^\mu \mbox{ such that }
\bar Z_\mu v^\mu =0.
\end{align*}
$\bar \Delta_{\mu\nu}$ is also the induced metric on the spacelike
hypersurface normal to $\bar Z^\mu$.
From the above point of view, $\bar n$, $\bar\epsilon$ and $\bar P$ are all
scalar observables which are independent of coordinate choices, but
dependent on the choice of observer.

In a static spacetime, an observer whose proper
velocity is proportional to $(\partial_t)^\mu$ is known as a static observer.
The choice of the static observer
$\bar{\mathcal{O}}$ leads to the following consequences:
(i) $\mathcal B^\mu$ and $\bar Z^\mu$ are colinear, $\mathcal B^\mu
= \beta \bar Z^\mu$;
(ii) $N^\mu$ has no spacial components with respect to $\bar Z^\mu$, i.e.
$N^\mu\bar\Delta_\mu{}^\nu = 0$;
(iii) the stress measured by $\bar O$ is isotropic.
The observer $\bar{\mathcal O}$ is actually an instantaneous
comoving observer, which means that the proper
velocity of the observer is identical to that of the fluid element.
Indeed, according to eqs.\eqref{Ncomp} and \eqref{Tcomp},
the particle number current $N^\mu$ and energy-momentum tensor $T^{\mu\nu}$
can be expressed in terms of $\bar Z^\mu$ as,
\begin{align}
N^\mu =\bar n \bar Z^\mu, \qquad
T^{\mu\nu} = \bar P \eta^{\mu\nu}
+ \frac{1}{c^2}\left(\bar \epsilon + \bar P\right) \bar Z^\mu \bar Z^\nu,
\label{NTZb}
\end{align}
which recovers the familiar result of the particle number current and
energy momentum tensor for a perfect fluid.

The parametrization of the particle number current and energy momentum tensor
using energy density and pressure is not the only choice. There exist other
sets of variables which fulfill the same purposes, some of which may even
be more preferable in some cases. For instance, when relativistic
transformations are under concern, it may be more reasonable to decompose
the above tensorial objects into irreducible parts which do not mix up
under local Lorentz boosts. For $T^{\mu\nu}$, one such decomposition is
given as follows,
\begin{align*}
&T^{\mu\nu} = \frac{1}{4} \mathcal{T} \eta^{\mu\nu}
+\frac{1}{4} \bar w \left(\bar \Delta^{\mu\nu}
+\frac{3}{c^2}\bar Z^\mu \bar Z^\nu\right), \\
&\mathcal{T}\equiv T^\sigma{}_\sigma = -
\frac{4\pi \re^{-\alpha}}{\lambda_{C}^{3} \zeta} mc^2 K_1(\zeta),
\quad \bar w \equiv \bar \epsilon + P
=  \frac{4\pi \re^{-\alpha}}{\lambda_{C}^{3} \zeta} mc^2 K_3(\zeta),
\end{align*}
where $\mathcal{T}$ is the trace of $T^{\mu\nu}$ and
$\bar w $ represents the enthalpy density. The decomposition of
$T^{\mu\nu}$ now consists in its trace and traceless parts.
Generally speaking, the trace part satisfies\footnote{Our convention
on the metric signature is $(-,+,+,+)$.} $\mathcal{T} \leq 0$,
and the enthalpy density satisfies
$\bar w \geq 0$.
Now we proceed to study the entropy density.
By use of eq.\eqref{NTZb}, the entropy current \eqref{Smu} becomes
\begin{align}
&S^\mu = \bar s \bar Z^\mu, \qquad
\bar s = k_{\mathrm{B}}\left(\beta\, \bar w + \bar n \alpha
\right).
\label{SZb}
\end{align}

Now we are in a position to analyze the
particle number density $n$, enthalpy density $w$
and entropy density $s$ measured by an arbitrary
instantaneous observer $\mathcal{O}$ with proper velocity $Z^\mu$.
Of course, $Z^\mu$ should still be a normalized timelike vector
at each instance. The normal projection tensor associated to the
observer $\mathcal{O}$ is naturally defined as
\[
\Delta^{\mu\nu}
= \eta^{\mu\nu}+\frac{1}{c^2} Z^\mu Z^\nu.
\]
Let us recall that, at the same spacetime event, any two
instantaneous observers can be connected by a (local) Lorentz boost.
To prove this statement, let us assume that $\bar Z^\mu$ and $Z^\mu$
are not proportional to each other. Otherwise, the two observers must
be identical. At the same event, we can always write
\begin{align}
\bar Z^\mu  = \gamma \left(Z^\mu + z^\mu\right), \quad
\text{where} \quad Z_\mu z^\mu = 0.
\label{boost}
\end{align}
Clearly, since $z^\mu$ is normal to $Z^\mu$, one has
\[
\Delta_{\mu\nu} Z^\mu=0,\qquad \Delta_{\mu\nu} z^\mu= z_\nu.
\]
From the normalization condition
for $\bar Z^\mu$ and $Z^\mu$, it is easy to see that
\[
\gamma=\frac{1}{\sqrt{1-z^2/c^2}},
\]
so $\gamma$ is nothing but the
Lorentz factor and $z^\mu$ can be regarded as the relative coordinate
time velocity between the two observers $\bar{\mathcal{O}}$ and $\mathcal{O}$
(notice that in a coordinate system in which the observer $\mathcal{O}$
remains static, $z^\mu$ has no temporal component and thus can be
considered as a 3-vector). Notice that although the
proper velocity of the observer has been changed, the Killing vector field
$\mathcal{B}^\mu$ appearing in the distribution function \eqref{eq1PDF}
remains untouched. Therefore,
$\mathcal{B}^\mu$ is not parallel to the proper velocity $Z^\mu$
of the new observer $\mathcal{O}$.

Inserting eq.\eqref{boost} into the expressions for $N^\mu, T^{\mu\nu}$
\eqref{NTZb} and $S^\mu$ \eqref{SZb},
we get
\begin{align}
&N^\mu = \gamma \bar n Z^\mu + \gamma \bar n z^\mu,
\nonumber\\
&T^{\mu\nu}
= \bar P \eta^{\mu\nu} + \frac{\gamma^2}{c^2}
\bar w \left[
Z^\mu Z^\nu + \left(Z^\mu z^\nu + Z^\nu z^\mu \right)
+ z^\mu z^\nu \right],
\label{tWz}\\
&S^\mu = \gamma \bar s Z^\mu + \gamma \bar s z^\mu. \nonumber
\end{align}
By definition, the particle number density measured by the observer $\mathcal{O}$
is
\begin{align}
n = - \frac{1}{c^2} Z_\mu N^\mu = \gamma \bar n,  \label{nnb}
\end{align}
which makes perfect
sense in terms of the length contraction effect and considering that
on the classical level the total number of particles in the system is
invariant. Similarly, the transformation for entropy density is
\begin{align}
s = - \frac{1}{c^2} Z_\mu S^\mu = \gamma \bar s, \label{ssb}
\end{align}
where again the $\gamma$ factor can be attributed to the length contraction
effect. In result, we can conclude that the total
entropy should be invariant under Lorentz boost after a volume
integration.

The definition for the pressure $P$ observed by the observer $\mathcal{O}$
contains some subtleties. We need first to separate
the energy momentum 4-vector $\mathcal P^\mu$
and the pure stress tensor $\mathcal{S}^{\mu\nu}$ from the energy momentum
tensor \eqref{tWz}:
\begin{align}
&\mathcal P^\mu \equiv -\frac{1}{c^2}Z_\rho T^{\rho\nu}\Delta_\nu{}^\mu
=\frac{\gamma^2\bar w}{c^2} z^\mu,\\
&\mathcal{S}^{\mu\nu}
\equiv T^{\rho\sigma}\Delta_\rho{}^\mu \Delta_\sigma{}^\nu
=\bar P \Delta^{\mu\nu}+ \frac{\gamma^2\bar w}{c^2} z^\mu z^\nu.
\end{align}
The existence of a spacelike momentum $\mathcal P^\mu$ reflects the fact that
the observer $\mathcal{O}$ is not comoving with the fluid. In other words,
$\mathcal P^\mu$ is purely a kinematic effect.
Notice also that, the second term in the expression for
$\mathcal{S}^{\mu\nu}$,
i.e.
\[
\tilde{\mathcal{S}}^{\mu\nu} \equiv \frac{\gamma^2\bar w}{c^2} z^\mu z^\nu
=\mathcal P^\mu z^\nu,
\]
is nothing but the kinematic momentum flow, and hence is also
purely kinematically originated, so it may be best referred to as
kinematic stress tensor. This part of the total stress tensor introduces
an anisotropy in the diagonal part, yielding different pressures in
different spatial directions. However, since the thermodynamic effects
has nothing to do with the global kinematics
of the fluid, it is better to define the thermodynamic pressure as the
trace of the difference between the total stress and the
kinematic stress divided by the dimension of the space.
In the present case, we have
\begin{align}
P =\frac{1}{3} \Delta_{\mu\nu}
(\mathcal{S}^{\mu\nu}-\tilde{\mathcal{S}}^{\mu\nu}) =\bar P,
\label{iso-P}
\end{align}
which indicates that the thermodynamic pressure is a relativistic invariant.

The definition of pressure is essential to determine the
correct enthalpy density, and therefore settles the transformation rules connecting
thermodynamic parameters measured by arbitrary observer $\mathcal O$
and comoving observer $\bar{\mathcal O}$. For enthalpy density,
equation \eqref{tWz} and the invariance of pressure \eqref{iso-P} imply
\begin{align}
w \equiv
\frac{1}{c^2} Z_\mu Z_\nu T^{\mu\nu} + P,
\qquad \quad
w = \gamma^2 \bar{w}. \label{wwb}
\end{align}
In view of the explicit expressions for particle density, entropy density, pressure and the
enthalpy density, we find that the local equilibrium state for arbitrary observer
can be parameterized by $\alpha$, $\beta$ and $\gamma$. And It is important to emphasize that
the thermodynamic pressure $P(\alpha, \beta)$ is isotropic and observer independent
whose derivative yields the following relations:
\begin{align*}
	n=-\gamma \beta \frac{\partial P}{\partial \alpha}, \quad w= -\gamma^2 \beta \frac{\partial P}{\partial \beta}, \quad s = -\gamma k_B \beta \left(\beta \frac{\partial P}{\partial \beta} + \alpha \frac{\partial P}{\partial \alpha}\right).
\end{align*}
Therefore, the total derivative of $P(\alpha,\beta)$ yields
\begin{align}
	-\gamma^{-1} s {\rm d} \left(\frac{1}{k_B \beta}\right) + {\rm d} P + \gamma^{-1}n {\rm d} \left(\frac{\alpha}{\beta}\right) = 0,
	\label{g-d}
\end{align}
for a comoving observer, i.e. when $\gamma =1$,
this equation is reminiscent to the local version of Gibbs-Duhem relation,
provided the parameters $\beta$ and $\alpha$ are respectively connected with
the inverse temperature and the chemical potential appropriately. In fact,
such a reminiscence is true, because a system
under detailed balance can be considered to be in local thermodynamic
equilibrium, and hence we can use the local Gibbs-Duhem relation to
identify the local temperature and chemical potential as
\begin{align}
\bar T = \frac{1}{k_{\mathrm{B}} \beta}, \qquad
\bar \mu = - \frac{\alpha}{\beta}.
\label{bTmu}
\end{align}
According to the basic principles of relativity, the choice of observers should not affect
physical identities obeyed by physical observables, although the value of each observable
may be affected by different choices. Among the thermodynamic relations,
in the light of the standpoint of Israel's theory \cite{Israel},
we now make our fundamental assumption that {\em the local Gibbs-Duhem relation
is invariant for different observers}.
In the present context, equation \eqref{g-d}
for different observers should correspond to the same physical identity.
In other words, \eqref{g-d} should be the invariant Gibbs-Duhem relation.
The direct consequences of the invariance of Gibbs-Duhem relation are the transformation rules of temperature and chemical potential:
\begin{align}
	T = \gamma^{-1} \frac{1}{k_B \beta} = \gamma^{-1} \bar{T}, \qquad \mu = -\gamma^{-1}\frac{\alpha}{\beta} = \gamma^{-1} \bar{\mu}.
	\label{TT}
\end{align}
It should be mentioned that,
although the temperature transformation given above has already been
suggested by de Broglie, Einstein and Planck, the one for chemical
potential is, to the best of our knowledge, a new result.
Finally, from eqs. \eqref{nnb}, \eqref{ssb}, \eqref{wwb}, \eqref{TT}
and the conventional definitions for Gibbs free energy and
Helmholtz free energy,
we find that the transformation rules for the densities of
all thermodynamic potentials can be expressed in an heuristic
unified form, $\varphi = \bar \varphi + z_\mu \mathcal P^\mu$.

Let us conclude this section by adding two extra remarks.

{\em Remark 1.} In relativistic physics, the spacetime dimension is
often treated as an adjustable parameter. Whenever one draws some conclusion in
relativistic physics, it is necessary to check the conclusion holds whether in
generic spacetime dimensions or in some specific dimension. On the other hand, the
behaviors of thermodynamic quantities are very sensitive to the dimension of the
underlying space. Therefore, it makes sense to check whether the transformation rules
uncovered in the present section is specific to 3+1 dimensional Minkowski spacetime or
they hold in arbitrary spacetime dimensions. In order to answer this question,
it is necessary to extend the formulation to arbitrary spacetime dimension
$(d+1)$. In this regard, it is important to note that the fluid configuration is
completely determined by $\left(\alpha,\mathcal B^\mu,n,w,\mathcal{T}\right)$,
wherein, for perfect fluid, $\alpha$ is constant, $\mathcal B^\mu$ is a
Killing vector field, and, in Minkowskian backgrounds (see Appendix),
\begin{align*}
&n =\gamma \left[\frac{2\,\re^{-\alpha}}{\lambda_C^d}
\left(\frac{2\pi}{\zeta}\right)^{\frac{d-1}{2}}\right]
K_{\frac{d+1}{2}}(\zeta) =\gamma \bar n, \\
&w = \gamma^2 mc^2 \left[\frac{2\,\re^{-\alpha}}{\lambda_C^d}
\left(\frac{2\pi}{\zeta}\right)^{\frac{d-1}{2}}\right]
K_{\frac{d+3}{2}}(\zeta) = \gamma^2 \bar w, \\
&\mathcal{T} = -mc^2 \left[\frac{2\, \re^{-\alpha}}{\lambda_C^d}
\left(\frac{2\pi}{\zeta}\right)^{\frac{d-1}{2}}\right]
K_{\frac{d-1}{2}}(\zeta).
\end{align*}
The exact results for the particle number density, the enthalpy density
and the trace of the energy momentum tensor
for perfect relativistic fluid in arbitrary spacetime
dimensions are, to our knowledge, not reported before in the literature.
Using these results, it will not be difficult to check that all the transformation rules
obtained in the present section are
independent of the spacetime dimension.

{\em Remark 2.} The temperature, chemical potential, particle number density,
entropy and enthalpy densities and the pressure of the perfect fluid are all
defined as observer-dependent scalars (or scalar densities). Their transformation rules
arise purely from the different choices of observers and have nothing to do with the
coordinate choices. It is not surprising that at the same spacetime event, any two
instantaneous observers can differ at most from each other by a local Lorentz boost
(which is {\em not} a coordinate transformation of the spacetime).
Such differences are independent of the choice of spacetime geometry. Therefore, it is
highly expected that the same transformation rules should hold in other spacetimes,
and we shall verify this expectation in Rindler spacetime in the next section.

\section{Perfect Rindler fluid and the area law of entropy}

The 1PDF under detailed balance given in eq.\eqref{eq1PDF} is valid not
only in Minkowski spacetime, but also in any spacetime admitting
a Killing vector field $\mathcal{B}_\mu$. In order to show the
influence of spacetime geometry on the description of perfect fluid,
let us move on to another familiar spacetime, i.e. the Rindler spacetime
with the line element
\begin{align}
\rd  s^2 = g_{\mu\nu}\rd x^\mu \rd x^\nu
= - \left(1+ \frac{\kappa x}{c^2}\right)^2 c^2 \rd t^2 +
\rd x^2 + \rd y^2 + \rd z^2.
\label{rindler}
\end{align}
The coordinates used in writing the line element \eqref{rindler} is
henceforth referred to as Rindler coordinates.
As is well known, the Rindler spacetime contains an accelerating horizon
which is located at $x_h = -c^2 \kappa^{-1}$ in the Rindler coordinate
system. In this spacetime, the static observer $\bar{\mathcal{O}}$ has
proper velocity $\bar Z^\mu =
\left(1+ \frac{\kappa x}{c^2}\right)^{-1} (\partial_t)^\mu $, which
is timelike but non-Killing.
The timelike Killing vector field $\mathcal B^\mu$ appearing in the 1PDF
\eqref{eq1PDF} normalized as $\mathcal B^2 = -\beta^2 c^2$ must be
proportional to $\bar Z^\mu$, i.e. $\mathcal B^\mu =\beta \bar Z^\mu
=\beta \left(1+ \frac{\kappa x}{c^2}\right)^{-1} (\partial_t)^\mu$,
which, together with the fact that $\bar Z^\mu$ is non-Killing, implies
that $\beta$ must not be constant, rather,
it has an explicit $x$-dependence,
\begin{align}
\beta(x) =- \beta_0 (x-x_h)/x_h,
\end{align}
where $\beta_0$ is a constant which equals to the value of
$\beta$ at $x=0$. Demanding that $\beta_0$ has a finite value
yields that $\beta$ becomes very small
as the system gets very close to the Rindler horizon.
On the other hand, if $\beta(x)$ approaches a finite
nonvanishing value as $x\to x_h$, then $\beta_0$ could
be infinitely large.

With the above choices of spacetime metric and the Killing vector field
$\mathcal B^\mu$, the particle number current $N^\mu$ and the energy
momentum tensor $T^{\mu\nu}$ can be explicitly calculated using
eq.\eqref{N&T}, yielding
\begin{align}
N^\mu &= \left[c\left(1+ \frac{\kappa x}{c^2}\right)^{-1}
\frac{4\pi \re^{-\alpha}}{\lambda_{C}^{3} \zeta} K_2(\zeta)
,0,0,0\right],    \label{Ncomp-R}  \\
T^{\mu \nu} &=\frac{4\pi \re^{-\alpha}}{\lambda_{C}^{3} \zeta}
mc^2 K_{2}(\zeta)
\left(\begin{array}{cccc}
\frac{\zeta K_{3}(\zeta)-K_2(\zeta)}{\left(1+ \frac{\kappa x}{c^2}\right)^2 \zeta K_{2}(\zeta) } & 0 & 0 & 0 \\
0 & \zeta^{-1} & 0 & 0 \\
0 & 0 & \zeta^{-1} & 0 \\
0 & 0 & 0 & \zeta^{-1}
\end{array}\right),
\label{Tcomp-R}
\end{align}
where $\zeta$ is still defined as $\zeta=\beta mc^2$ and hence is
non-constant.

Unlike the case of Minkowski spacetime, now we should not expect to
read off the particle number density $\bar n$, the energy
density $\bar \epsilon$ and the pressure
$\bar P$ directly from the appropriate components of
$N^\mu$ and $T^{\mu\nu}$ respectively as we did in eqs.\eqref{nb}, \eqref{Pb}
and \eqref{eb}. Rather, we should think of these objects as scalar densities
defined in eq.\eqref{covdefs} and \eqref{bPstress}.
Without much effort it can be shown that $\bar n$, $\bar \epsilon$ and
$\bar P$ for the Rindler fluid are given respectively by the
following expressions,
\begin{align}
&\bar n = - \frac{1}{c^2} \bar Z_\mu N^\mu
= \frac{4\pi \re^{-\alpha}}{\lambda_{C}^{3} \zeta} K_{2}(\zeta),
\label{nbar-R}\\
&\bar \epsilon = \frac{1}{c^2} \bar Z_\mu \bar Z_\nu T^{\mu\nu}
= \frac{4\pi \re^{-\alpha}}{\lambda_{C}^{3} \zeta}
mc^2 \left[K_{3}(\zeta)-\zeta^{-1}K_{2}(\zeta)\right], \label{ebar-R}\\
&\bar P = \frac{1}{3} \bar \Delta_{\mu\nu} \bar{\mathcal{S}}^{\mu\nu}
= \frac{4\pi \re^{-\alpha}}{\lambda_{C}^{3} \zeta^2} mc^2 K_{2}(\zeta)
=\frac{\bar n}{\beta},
\label{Pbar-R}
\end{align}
and the decomposition like \eqref{NTZb} for $N^{\mu}$
and $T^{\mu\nu}$ still
holds, while the only difference comes from the different choice of
spacetime metric:
\begin{align}
N^\mu =\bar n \bar Z^\mu, \qquad
T^{\mu\nu} = \bar P g^{\mu\nu}
+ \frac{1}{c^2}\left(\bar \epsilon + \bar P\right) \bar Z^\mu \bar Z^\nu.
\label{NTZbR}
\end{align}
These results indicate that the static observer $\bar{\mathcal{O}}$ is
instantaneously comoving with the fluid.
Moreover, the entropy density $\bar s$ can be evaluated
explicitly,
\begin{align}
\bar s &= - \frac{1}{c^2} \bar Z_\mu S^\mu
= - \frac{k_{\mathrm{B}}}{c^2} \bar Z_\mu \left[ \left(1+\alpha\right)N^\mu
- \mathcal B_\nu T^{\mu\nu}\right]\nonumber\\
&= k_{\mathrm{B}} \frac{4\pi \re^{-\alpha}}{\lambda_{C}^{3} }
\left[K_3 (\zeta) + \alpha \frac{K_{2}(\zeta)}{\zeta}\right]\nonumber\\
&= k_{\mathrm{B}}\left(\beta\, \bar w + \bar n \alpha \right).
\label{sbar-R}
\end{align}
All the results given in eqs.\eqref{nbar-R}-\eqref{sbar-R} are in perfect
agreement with their Minkowski analogues, as shown in
eqs.\eqref{nb}-\eqref{Pb}, \eqref{NTZb} and \eqref{SZb}.
If we proceed to an arbitrary Rindler observer, the
same transformation rules for $T,\mu,P,s$ and $w$ will be recovered,
with the same interpretations for the parameters $\alpha, \beta$.
The $x$ dependence of $\beta$ is then attributed to the
Tolman-Ehrenfest effect in Rindler spacetime.
We will not bother with details on the transformation rules
in Rindler spacetime but only mention that, in this case,
the relation \eqref{boost} between the proper velocities of
different observers results in a position dependent boost
factor $\gamma$, because now $z^2 = g_{\mu\nu}z^\mu
z^\nu$ is position dependent. Such boosts must be considered to be
performed in the tangent space of the spacetime at each event, and are
known as local Lorentz boosts.

One thing to be noted, however, is that all the transformation rules
mentioned above reflect only the behaviors of the corresponding quantities
measured at the same spacetime event by different observers. If one wishes
to compare the same quantities at different spacetime events, or add up
some of the local densities to get the corresponding total quantities, things
would become drastically different between Minkowski and Rindler cases.
For the Minkowski case, nothing special needs to be taken care of, because
the inverse temperature $\beta$ is constant and all the local densities are
uniform. For the Rindler case, however, the first thing to be noted is
the non-constancy of the inverse temperature $\beta$.
At different spatial locations with $x$ coordinates $x_1$
and $x_2$, one has
\begin{align*}
\frac{\zeta_1}{\zeta_2} = \frac{\beta_1}{\beta_2}
= \frac{x_1 - x_h}{x_2 - x_h}.
\end{align*}
This is the well known Tolman-Ehrenfest effect \cite{Ehr}
and it implies that the local densities \eqref{nbar-R}, \eqref{ebar-R},
\eqref{sbar-R} as well as the pressure \eqref{Pbar-R} in Rindler spacetime
are all non-uniform. Due to the constancy of the parameter $\alpha$,
the chemical potential $\bar \mu$ in Rindler spacetime is also
non-uniform.

Now let us consider a box of gas, of which the height and the bottom area
are respectively $L$ and $A$, and orient the box such that the bottom is
parallel to the horizon. Let the coordinate distance of the bottom of
the box to the horizon be $\delta$. The relativistic coldness at the bottom
of the box is denoted by $\zeta_{\mathrm{bot}}$,
then at any height within the box we have
$\zeta = (x-x_h)\zeta_{\mathrm{bot}}/\delta$,
and at the top of the box we have $\zeta_{\mathrm{top}}
= (L+\delta) \zeta_{\mathrm{bot}}/\delta$.

The total number of particles and the total entropy in the box
can be calculated by a direct spatial volume integration
of \eqref{nbar-R} and \eqref{sbar-R}, yielding
\begin{align*}
\bar N &= A \int_{x_h + \delta}^{x_h + \delta+ L} \bar n(x) \rd x
= A\frac{4\pi \re^{-\alpha}}{\lambda_{C}^{3}}
\frac{\delta}{\zeta_{\mathrm{bot}}}
\left[\frac{K_1(\zeta_{\mathrm{bot}})}{\zeta_{\mathrm{bot}}} - \frac{K_1(\zeta_{\mathrm{top}})}{\zeta_{\mathrm{top}}}\right], \\
\bar S &=A \int_{x_h + \delta}^{x_h + \delta+ L} \bar s(x) \rd x
= A\frac{4\pi k_{\mathrm{B}} \re^{-\alpha}}{\lambda_{C}^{3}}
\frac{\delta}{\zeta_{\mathrm{bot}}}
\left[K_2(\zeta_{\mathrm{bot}}) - K_2(\zeta_{\mathrm{top}}) \right]
+(\alpha+2) k_{\mathrm{B}} \bar N.
\end{align*}
\noindent When the bottom of the box is very close to the horizon,
e.g. $\delta \ll L$, the coldness at the bottom and top of the box
behave as $\zeta_{\mathrm{bot}} \ll \zeta_{\mathrm{top}} $. Recalling
that both $K_1(\zeta)$ and $K_2(\zeta)$ are monotonically decreasing
functions, we have $K_1(\zeta_{\mathrm{bot}}) \gg K_1(\zeta_{\mathrm{top}})$,
$K_2(\zeta_{\mathrm{bot}}) \gg K_2(\zeta_{\mathrm{top}})$.
Then from eq.\eqref{nbar-R} one finds that
the particles tend to be gathered at the bottom of the box,
and the total number of particles and entropy can be approximated as
\begin{align*}
\bar N &\approx A  \frac{4\pi \re^{-\alpha}}{\lambda_{C}^{3}}
\frac{\delta}{\zeta_{\mathrm{bot}}^2}
K_1(\zeta_{\mathrm{bot}}),\\
\bar S &\approx A
\frac{4\pi k_{\mathrm{B}}\re^{-\alpha}}{\lambda_{C}^{3}}
\frac{\delta}{\zeta_{\mathrm{bot}}^2}
\left[(\alpha +2)K_1(\zeta_{\mathrm{bot}})
+ \zeta_{\mathrm{bot}} K_2(\zeta_{\mathrm{bot}})\right],
\end{align*}
which are both independent of the hight of the box and are
proportional to the area of  the bottom of the box.
These results are similar to those in
\cite{Padmanabhan2011,Padmanabhan2017,Kim2017} where the particle number and
entropy are obtained through completely different approaches.
For closed classical systems, the number of particles needs to be constant,
and the expression for $\bar N$ actually determines the chemical potential
distribution implicitly.
Comparing the values of $\bar N$ and $\bar S$ in the above case, we get
\begin{align}
\bar S \approx \bar N k_{\mathrm{B}}\left[(\alpha+2)
+\frac{\zeta_{\mathrm{bot}} K_2(\zeta_{\mathrm{bot}})}
{K_1(\zeta_{\mathrm{bot}})}\right],
\label{addit}
\end{align}
wherein the factor besides $\bar N k_{\mathrm{B}}$ depends purely on the
coldness at the bottom of the box but not on any other physical parameters because
of the constancy of $\alpha$. The relationship \eqref{addit}
indicates that the area dependence of the entropy does not
break its additivity.

\section{Refined Saha equation}

As we have shown in the last two sections, the local particle
number density $\bar n$ takes the same form in both Minkowski and
Rindler spacetimes.  By means of the explicit expressions
for $\bar n$, see eqs. \eqref{nb} or \eqref{nbar-R}, we find that
the fugacity can be written as
\begin{align}
\mathfrak f = \re^{-\alpha-\zeta} = \bar n \lambda_T^3,
\qquad
\lambda_T = \frac{\lambda_C}{\sqrt[3]{4\pi
		\re^\zeta \zeta^{-1} K_2(\zeta)}}.
\label{AL}
\end{align}

To understand the
physical meaning of $\lambda_T$, we now focus on two extremal cases,
$\zeta\gg1$ and $\zeta\ll1$. The former choice corresponds to
the non-relativistic limit which applies to baryons, and the latter choice
corresponds to the ultra-relativistic limit, which is attained if
the temperature is ultra high or the rest mass is very small. It is easy to check that,
in the non-relativistic limit and ultra-relativistic limit,
$\lambda_T$ coincides respectively with the thermal de Broglie wavelength
for massive and massless particles
\begin{align}
&\lambda_T \approx \lambda_{\text{NR}}=\frac{h}{\sqrt{2 \pi m k_{\mathrm{B}} \bar T}}, ~~~
\left(\zeta\gg1\right),
\label{nonrlT}\\
&\lambda_T \approx \lambda_{\text{UR}}=\frac{c h}{2 \pi^{1 / 3} k_{\mathrm{B}}\bar T}, ~~~~~
\left(\zeta\ll1\right).
\end{align}
In this sense, $\lambda_T$ may be referred to as the relativistic
thermal de Broglie wavelength. It is worth mentioning that, at generic coldness $\zeta$,
the relativistic thermal wavelength $\lambda_T$ can be significantly different from
either $\lambda_{\text{NR}}$ or $\lambda_{\text{UR}}$, as can be seen in Fig.1.

\begin{figure}[h]
	\begin{center}
		\includegraphics[height=.3\textheight]{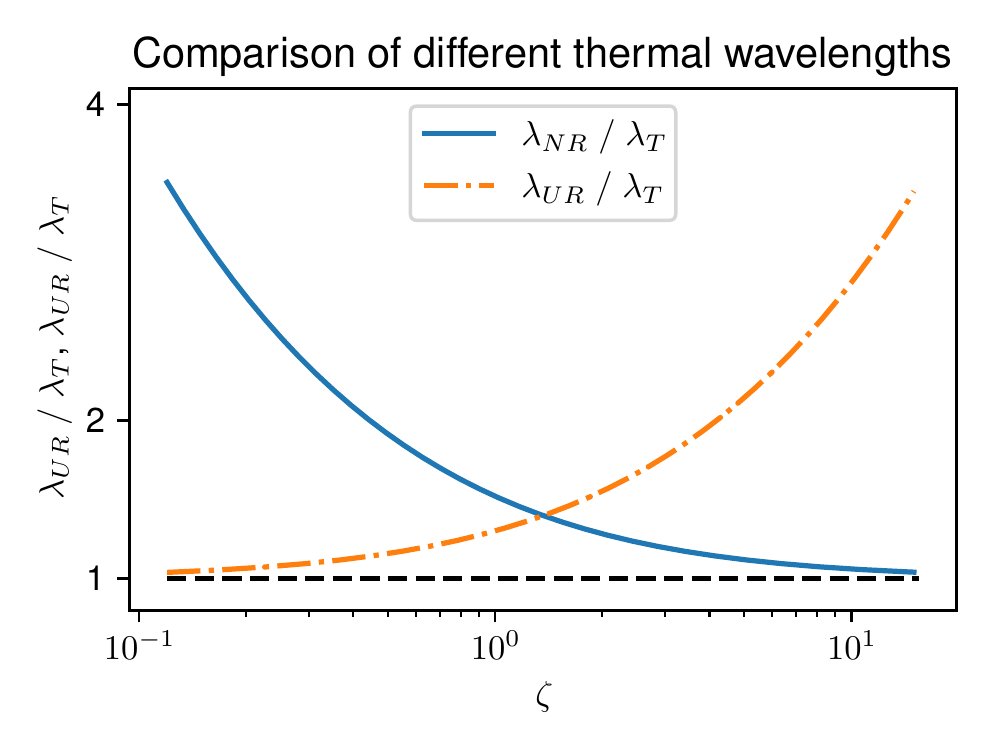}	
		
		\caption{Comparison between $\lambda_T$ and $\lambda_{\text{NR}}, \lambda_{\text{UR}}$}
	\end{center}
\end{figure}

By use of the relativistic thermal de Broglie wavelength $\lambda_T$ ,
we can recast the expressions for $\bar n$ in the form
\begin{align*}
\bar n= \frac{1}{(\lambda_T)^3}\re^{-\alpha-\zeta}
=\frac{1}{(\lambda_T)^3}\re^{(\bar\mu-mc^2)/k_{\mathrm{B}}\bar T}.
\end{align*}
If the particle under consideration has some intrinsic quantum number,
then the above expression needs to be multiplied by the corresponding
intrinsic quantum degeneracy $\mathfrak{g}$, i.e.
\begin{align}
\bar n= \frac{\mathfrak{g}}{(\lambda_T)^3}\re^{-\alpha-\zeta}
=\frac{\mathfrak{g}}{(\lambda_T)^3}\re^{(\bar\mu-mc^2)/k_{\mathrm{B}}\bar T}.
\label{nexp}
\end{align}
This expression allows us to get a refined version of the
famous Saha equation \cite{saha} which is related to the local ``chemical
equilibrium'' for systems containing several different species of
particles among which chemical reactions (or ionizations/recombinations)
may take place. Let us assume that the system contains 3
different particle species $A, B, C$ among which $C=AB$ is the composite
particle may come as the result of the chemical reaction
\[
A+B \rightleftharpoons C.
\]
Assume also that the interaction between different particles is
negligible unless they collide and/or make
the above chemical reaction. Under the above assumptions, the chemical
potentials for different particle species satisfy
the condition $\bar\mu_A+\bar\mu_B=\bar\mu_C$ if
chemical reaction reaches a local equilibrium.
Therefore, it follows from eq.\eqref{nexp} that
\begin{align}
\frac{\bar n_A \bar n_B}{\bar n_C}
&=\frac{\mathfrak{g}_A\mathfrak{g}_B}{\mathfrak{g}_C}
\left(\frac{\lambda_T^{(C)}}
{\lambda_T^{(A)}\lambda_T^{(B)}}\right)^3
\re^{-\Delta E/k_{\mathrm{B}}\bar T},
\label{RSaha}
\end{align}
where $\Delta E=(m_A+m_B-m_C)c^2$ is the binding energy of the
composite particle $C$. If $\lambda_T$ takes the form of
eq.\eqref{nonrlT}, then eq.\eqref{RSaha} is precisely the famous
Saha equation. However, as we have shown in Section \ref{sec:Minkowski},
$\lambda_T$ actually takes the form \eqref{AL}, which is different
from \eqref{nonrlT} unless $\zeta\gg1$.
Therefore, at smaller coldness (or higher temperature),
eq.\eqref{RSaha} gives a relativistically refined version of
the standard Saha equation. For gaseous systems with a single
species of atoms which have several ionized states,
eq.\eqref{RSaha} reduces into
\[
\frac{\bar n_{i+1} \bar n_e}{\bar n_i}
=2\frac{\mathfrak{g}_{i+1}}{\mathfrak{g}_i}
\frac{1}{\left(\lambda_T^{(e)}\right)^3}
\re^{-\Delta E_{i+1}/k_{\mathrm{B}}\bar T},
\]
where $\bar n_{i}$ is the number density of the atom in the $i^{th}$
ionized state, $\Delta E_{i+1}$ is the energy required to remove the
$(i+1)^{th}$ electron, and $\lambda_T^{(e)}$ is the
relativistic thermal de Broglie wavelength for the electron.

\section{Concluding remarks}

Relativistic kinetic theory is a powerful tool for analyzing
macroscopic behaviors for systems undergoing relativistic motion.
In this work, using relativistic kinetic theory,
we considered the long standing problem on
the relativistic transformation rules for some basic thermodynamic
quantities, including the temperature $T$ and the chemical potential $\mu$.
Our the major results are listed as follows:
\[
T = \gamma^{-1} \bar T, \quad s=\gamma \bar s, \quad
\mu = \gamma^{-1} \bar\mu, \quad n = \gamma \bar n, \quad
P = \bar P, \quad w = \gamma^2 \bar w,
\]
wherein $\gamma$ is the (local) Lorentz factor, and
quantities with/without a bar are respectively measured by the
instantaneous comoving observer $\bar{\mathcal{O}}$ and an
arbitrary instantaneous observer $\mathcal{O}$, which are
interrelated by a (local) Lorentz boost.
These transformation rules, supplemented with
the well acknowledged contraction rule $\delta v = \gamma^{-1} \delta \bar v$ 
for spatial volume element $\delta v$,
constitute a complete set of transformation rules for all
thermodynamic parameters which
hold in both Minkowski and Rindler spacetimes, and we do not see any
reason why they would not hold in other spacetimes.
The transformation
rule for $T$ suggests that a moving body appears to be colder,
supporting the first view by de Broglie, Einstein and Planck \emph{et al}.
Moreover, our study adds some novel elements to the existing transformation rules,
for instance, the rules for
chemical potential and enthalpy density have not been reported elsewhere.

In the case of perfect Rindler fluid, if we place the fluid in a box whose
bottom is parallel to the Rindler horizon and let the bottom be
located sufficiently close to the horizon, then our calculation shows
that the total entropy and the total number of particles are
proportional to the area of the bottom of the box, and both are
independent of the height of the box. By extending to
spacetimes with event horizons, similar analysis might help
to understand the area law of black hole entropy.

Finally, the exact results for the particle number density allow us
to get a relativistically refined version of the famous Saha equation.
At extremely high temperatures, or in the ultra relativistic limits,
the refinement could be significant, as in these cases the
relativistic thermal de Broglie wavelength \eqref{AL} could be
significantly different from its non-relativistic counterpart \eqref{nonrlT}.

\section*{Acknowledgement}

This work is supported by the National Natural Science Foundation of
China under the grant No. 11575088, Hebei NSF under grant No. A2021205037
and the fund of Hebei Normal University under grant No.
L2020B04. XH would like to thank Shao-Jiang Wang
and Bin Wu for useful discussions.

\section*{Appendix}

In the literature on relativistic kinetic theory,
the expressions of local equilibrium quantities are mostly presented in
a concrete spacetime dimension, mostly taken to be 4. In case that
one may be interested in physical rules in generic spacetime dimensions,
we now present the detailed calculations for all the macroscopic densities
used in the main text in $(d+1)$-dimensional Minkowski spacetime.

Let us start with the cartesian coordinates and take the Killing
vector field $\mathcal B^\mu$ to be
$\mathcal B^\mu = \beta(\partial_t)^\mu =\beta \bar Z^\mu$.
In this coordinates we can always parametrize $p^\mu$ and $Z^\mu$ as
\begin{align*}
p^\mu &= mc(\cosh\vartheta, ~\sinh\vartheta\cos\theta_{d-1},
~\sinh\vartheta\sin\theta_{d-1}\cos\theta_{d-2}, \cdots,\\
&\qquad\qquad
~\sinh\vartheta\sin\theta_{d-1}\cdots\sin\theta_{2}\cos\theta_1,
~\sinh\vartheta\sin\theta_{d-1}\cdots\sin\theta_{2}\sin\theta_1), \\
Z^\mu &= c(\cosh\eta, ~\sinh\eta\cos\varTheta_{d-1},
~\sinh\eta\sin\varTheta_{d-1}\cos\varTheta_{d-2}, \cdots\\
&\qquad\qquad
~\sinh\vartheta\sin\varTheta_{d-1}\cdots\sin\varTheta_{2}\cos\varTheta_1,
~\sinh\vartheta\sin\varTheta_{d-1}\cdots\sin\varTheta_{2}\sin\varTheta_1),
\end{align*}
where $\theta_1,\varTheta_1\in[0,2\pi)$, and
$\theta_i,\varTheta_i\in[0,\pi)$ for $i\geq2$. To make the above parametrization
also work for $d=1$, we also introduce $\theta_0=\varTheta_0=0$.

It is straightforward to check that $p^\mu$ and $Z^\mu$ parametrized as above
automatically satisfy the on-shell conditions $p^2 = -m^2 c^2$,
$Z^2 = - c^2$, and
\[
|\vec p| = \sqrt{\left(p^1\right)^2
+ \left(p^2\right)^2
+\cdots + \left(p^d\right)^2} = mc \sinh \vartheta,\quad
\mathrm d |\vec p| = p^0 \mathrm d \vartheta
= |p_0| \mathrm d \vartheta,
\]
so the momentum space volume element reads
\[
\varpi=\frac{\mathrm d^dp}{|p_0|} = \frac{|\vec p|^{d-1}
\mathrm d |\vec p| \, \mathrm d \Omega_{d-1}}{|p_0|} =
(mc\sinh\vartheta)^{d-1}
\mathrm d \vartheta \, \mathrm d \Omega_{d-1},
\]
where for $d=1$ $\mathrm d \Omega_{0} = 1$, and for $d\geq2$,
$\mathrm d \Omega_{d-1}
= \sin^{d-2} \theta_{d-1} \mathrm d \theta_{d-1}
\mathrm d \Omega_{d-2}$. The coordinate choice freedom allows us to set
$\varTheta_{d-1}=0$, which will greatly simplify the forthcoming calculations.
Moreover, comparing $\bar Z^\mu$ with $Z^\mu$ and considering the
relation \eqref{boost}, one recognizes that $\gamma=\cosh\eta$ is precisely
the Lorentz factor related to the boost between the observers $\bar{\mathcal O}$
and $\mathcal O$.

Before delving into the analysis, we recall the following three integral
representations for the modified Bessel function of the second kind,
which are frequently used in our calculations,
\begin{align}
K_{\nu}(\zeta)
&= \int^\infty_0 \re^{-\zeta \cosh\vartheta}
\cosh(\nu\vartheta)\mathrm d\vartheta, \\
K_{\nu}(\zeta)
&=\left(\frac{\zeta}{2}\right)^\nu
\frac{\sqrt{\pi}}{\Gamma\left(\nu+\frac{1}{2}\right)}
\int^\infty_0 \sinh^{2\nu} \vartheta\,
\re^{-\zeta \cosh \vartheta} \rd \vartheta , \label{Bessel2}\\
K_{\nu}(\zeta)&=\left(\frac{\zeta}{2}\right)^{\nu} \frac{\Gamma\left(\frac{1}{2}\right)}{\Gamma\left(n+\frac{1}{2}\right)}
\int_{1}^{\infty} \re^{-\zeta \vartheta}
\left(\vartheta^{2}-1\right)^{\nu-\frac{1}{2}} \rd \vartheta.
\label{Bessel3}
\end{align}
From the integral representation \eqref{Bessel2}, we can get
\[
\left(\frac{\rd}{\rd \zeta}\right)^{l}
\left[\frac{K_\nu(\zeta)}{\zeta^\nu}\right]
= \frac{(-1)^l\sqrt{\pi}}{2^\nu\,\Gamma
\left(\nu+\frac{1}{2}\right)}
\int^\infty_0 \rd \vartheta \sinh^{2\nu} \vartheta
\cosh^l \vartheta\,
\re^{-\zeta \cosh \vartheta}.
\]
On the other hand, from the representation \eqref{Bessel3} we
obtain the recurrence relations
\[
\left(\frac{\rd}{\zeta\rd \zeta}\right)^l
\left[\frac{K_\nu(\zeta)}{\zeta^\nu}\right]
= (-1)^l\frac{K_{\nu+l}(\zeta)}{\zeta^{\nu+l}}.
\]
It follows that
\begin{align*}
&\int^\infty_0 \rd \vartheta \sinh^{2\nu} \vartheta
\,\re^{-\zeta \cosh \vartheta}
= \frac{1}{\sqrt{\pi}}
\left(\frac{2}{\zeta}\right)^\nu
\Gamma\left(\nu+\frac{1}{2}\right) K_{\nu}(\zeta), \\
&\int^\infty_0 \rd \vartheta \sinh^{2\nu} \vartheta
\cosh \vartheta\,\re^{-\zeta \cosh \vartheta}
= \frac{1}{\sqrt{\pi}}
\left(\frac{2}{\zeta}\right)^\nu
\Gamma\left(\nu+\frac{1}{2}\right) K_{\nu+1}(\zeta), \\
&\int_0^\infty \rd \vartheta \sinh^{2\nu} \vartheta
\cosh^2 \vartheta\,\re^{-\zeta \cosh \vartheta}
= \frac{1}{\sqrt{\pi}} \left(\frac{2}{\zeta}\right)^\nu
\Gamma\left(\nu+\frac{1}{2}\right)
\left[K_{\nu+2}(\zeta)-\zeta^{-1}K_{\nu+1}\right].
\end{align*}
These integrations are sufficient for us to analyze the following
observable quantities,
\begin{align}
n &= - \frac{1}{c^2} Z_\mu N^\mu
= -\frac{\re^{-\alpha}}{ch^d}
\int \frac{\mathrm d^d p}{|p_0|}
Z_\mu p^\mu \re^{\mathcal B_\mu p^\mu},
\label{n-ddim}\\
\epsilon &= \frac{1}{c^2} Z_\mu Z_\nu T^{\mu\nu}
= \frac{\re^{-\alpha}}{ch^d}
\int \frac{\mathrm d^d p}{|p_0|}
\left(Z_\mu p^\mu\right)^2 \re^{\mathcal B_\mu p^\mu}, \\
\mathcal T &= \frac{c\,\re^{-\alpha}}{h^d}
\int \frac{\sqrt g\,\mathrm d^dp}{|p_0|} p^2
\re^{\mathcal B_\rho p^\rho}
= - \frac{m^2 c^3 \re^{-\alpha}}{h^d}
\int \frac{\mathrm d^dp}{|p_0|} e^{\mathcal B_\mu p^\mu},
\label{T-ddim}
\end{align}
where $\mathcal B_\mu p^\mu = \zeta \cosh \vartheta$ where $\zeta$ is
the relativistic coldness, and the contraction
$Z_\mu p^\mu$ is simplified after taking $\varTheta_{d-1} = 0$,
\[
Z_\mu p^\mu = mc^2\left(\sinh\eta \sinh\vartheta \cos\theta_{d-1}
- \cosh\eta \cosh\vartheta\right)= mc^2 \mathcal
C(\eta,\vartheta,\theta_{d-1}).
\]

When $d=1$ there is only one spacial direction and
the above integrations can be carried out with ease.
First, the trace of energy-momentum tensor reads
\begin{align}
\mathcal T
&= -  \frac{ \re^{-\alpha}}{\lambda_C} mc^2
\int_{-\infty}^{+\infty} \mathrm d \vartheta
\re^{- \zeta \cosh\vartheta}
= - \frac{ 2 \re^{-\alpha}}{\lambda_C} mc^2 K_0(\zeta).
\label{T:d=1}
\end{align}
For the particle number density and energy density, after a few lines of simple
calculations we obtain
\begin{align}
n &= - \frac{\re^{-\alpha}}{\lambda_C} \int_{-\infty}^{\infty} \mathrm d \vartheta
\left(\sinh\eta \sinh\vartheta
- \cosh\eta \cosh\vartheta\right) \re^{- \zeta \cosh\vartheta} \nonumber \\
&= \frac{2\re^{-\alpha}}{\lambda_C} \cosh\eta  \int_0^\infty \mathrm d \vartheta
\cosh\vartheta \re^{- \zeta \cosh\vartheta} \nonumber \\
&= \gamma \frac{2\re^{-\alpha}}{\lambda_C} K_1(\zeta),
\label{n:d=1}
\end{align}
and
\begin{align}
\epsilon
&= \frac{\re^{-\alpha}}{\lambda_C} mc^2 \int_{-\infty}^{\infty} \mathrm d \vartheta
\left(\sinh\eta \sinh\vartheta
- \cosh\eta \cosh\vartheta\right)^2 e^{- \zeta \cosh\vartheta} \nonumber \\
&= \frac{2\re^{-\alpha}}{\lambda_C} mc^2 \int_0^\infty \mathrm d \vartheta
\left[\gamma^2 \cosh2\vartheta - \sinh^2\vartheta \right]
e^{- \zeta \cosh\vartheta} \nonumber \\
&= \frac{2\re^{-\alpha}}{\lambda_C} mc^2
\left[\gamma^2 K_2(\zeta) - \frac{K_1(\zeta)}{\zeta} \right].
\label{e:d=1}
\end{align}
When $d \geq 2$, the integrations \eqref{n-ddim}-\eqref{T-ddim}
can be evaluated in a unified way,
\begin{align}
n &= - \frac{\re^{-\alpha}}{\lambda_C^d}
\int \rd \Omega_{d-1}
\int_0^\infty \mathrm d \vartheta
\sinh^{d-1} \vartheta \, \mathcal C (\eta,\vartheta,\theta_{d-1})
\re^{- \zeta \cosh\vartheta}  \nonumber \\
&= \frac{\re^{-\alpha}}{\lambda_C^d}
\cosh\eta \int \rd \Omega_{d-1}
\int_0^\infty \mathrm d \vartheta
\sinh^{d-1} \vartheta \cosh\vartheta\,\re^{- \zeta \cosh\vartheta} \nonumber \\
&= \frac{\mathcal A_{d-1} \re^{-\alpha}}{\lambda_C^d}
\gamma \int_0^\infty \mathrm d \vartheta
\sinh^{d-1} \vartheta \cosh\vartheta\,\re^{- \zeta \cosh\vartheta} \nonumber \\
&=\gamma\frac{2\, \re^{-\alpha}}{\lambda_C^d}
\left(\frac{2\pi}{\zeta}\right)^{\frac{d-1}{2}}
K_{\frac{d+1}{2}}(\zeta),
\label{n:d>1}
\end{align}
where $\displaystyle\mathcal A_{d-1} =\int \rd \Omega_{d-1}
= \frac{2\pi^{d/2}}{\Gamma \left(\frac{d}{2}\right)}$ is the
area of a $(d-1)$-dimensional unit sphere,
\begin{align}
\epsilon
&= \frac{\re^{-\alpha}}{\lambda_C^d} mc^2
\int \rd \Omega_{d-1} \int_0^\infty \mathrm d \vartheta
\sinh^{d-1} \vartheta\,
\mathcal C^2 (\eta,\vartheta,\theta_{d-1})
\, \re^{- \zeta \cosh\vartheta} \nonumber \\
&= \frac{\re^{-\alpha}}{\lambda_C^d}
mc^2 \left[\gamma^2 \mathcal{A}_{d-1}
\int_0^\infty \mathrm d \vartheta
\sinh^{d-1}\vartheta \cosh^2\vartheta
\,\re^{- \zeta \cosh\vartheta} \right. \nonumber \\
& \qquad \qquad \qquad
\left. + (\gamma^2-1) \mathcal{A}_{d-2} \frac{\sqrt{\pi}}{2}
\frac{\Gamma\left(\frac{d-1}{2}\right)}{\Gamma\left(\frac{d}{2}+1\right)}
\int_0^\infty \mathrm d \vartheta \sinh^{d+1}\vartheta
\,\re^{- \zeta \cosh\vartheta} \right] \nonumber \\
&= \frac{\re^{-\alpha}}{\lambda_C^d} mc^2 \mathcal{A}_{d-1}
\int_0^\infty \mathrm d \vartheta \left[\gamma^2
\sinh^{d-1}\vartheta \cosh^2\vartheta
+\frac{\gamma^2-1}{d} \sinh^{d+1}\vartheta \right]
\re^{-\zeta \cosh\vartheta}  \nonumber \\
&= \frac{\re^{-\alpha}}{\lambda_C^d}
\frac{\mathcal{A}_{d-1}}{\sqrt{\pi}}
\left(\frac{2}{\zeta}\right)^{\frac{d-1}{2}}
\Gamma\left(\frac{d}{2}\right) mc^2
\left[\gamma^2 \left(K_{\frac{d+3}{2}}-\zeta^{-1}K_{\frac{d+1}{2}}\right)
+\frac{\gamma^2-1}{d}\frac{2}{\zeta}\frac{d}{2}K_{\frac{d+1}{2}} \right] \nonumber \\
&= \frac{2\,\re^{-\alpha}}{\lambda_C^d}
\left(\frac{2\pi}{\zeta}\right)^{\frac{d-1}{2}}
mc^2 \left[\gamma^2 K_{\frac{d+3}{2}}(\zeta)
- \zeta^{-1} K_{\frac{d+1}{2}}(\zeta)\right],
\label{e:d>1}
\end{align}
and
\begin{align}
\mathcal T
&= -  \frac{\re^{-\alpha}}{\lambda_C^d} \mathcal{A}_{d-1} mc^2
\int_0^\infty \mathrm d \vartheta
\sinh^{d-1}\vartheta \,\re^{- \zeta \cosh\vartheta} \nonumber \\
&= -\frac{\re^{-\alpha}}{\lambda_C^d}
\frac{\mathcal{A}_{d-1}}{\sqrt{\pi}}
\left(\frac{2}{\zeta}\right)^{\frac{d-1}{2}}
\Gamma\left(\frac{d}{2}\right) mc^2  K_{\frac{d-1}{2}}(\zeta) \nonumber \\
&= -\frac{2\, \re^{-\alpha}}{\lambda_C^d}
\left(\frac{2\pi}{\zeta}\right)^{\frac{d-1}{2}}
mc^2 K_{\frac{d-1}{2}}(\zeta).
\label{T:d>1}
\end{align}
Interestingly, the results \eqref{n:d>1}-\eqref{T:d>1} calculated for $d \geq 2$
are also valid for $d=1$, as we have shown in eqs.\eqref{T:d=1}-\eqref{e:d=1} .

For the thermodynamic pressure we refer to \eqref{iso-P},
with the replacement $\displaystyle\frac{1}{3}\to\frac{1}{d}$.
By evaluating the isotropic pressure in the rest frame where
$\gamma = 1$, we get
\begin{align}
P = \bar P = \frac{\mathcal T + \bar \epsilon}{d} =
\frac{2\,\re^{-\alpha}}{\lambda_C^d}
\left(\frac{2\pi}{\zeta}\right)^{\frac{d-1}{2}}
\zeta^{-1} mc^2 K_{\frac{d+1}{2}}(\zeta),
\label{POA}
\end{align}
finally the enthalpy density is obtained to be
\begin{align}
w = \epsilon + P
= \gamma^2 \left[\frac{2\,\re^{-\alpha}}{\lambda_C^d}
\left(\frac{2\pi}{\zeta}\right)^{\frac{d-1}{2}}\right]
mc^2  K_{\frac{d+3}{2}}(\zeta). \label{wOA}
\end{align}
This finishes our calculations for the relevant quantities in arbitrary dimensions.

\providecommand{\href}[2]{#2}\begingroup\raggedright\endgroup

\end{document}